\documentclass[runningheads,citeauthoryear]{apinv0}
\usepackage{epsfig,cite,graphics}
\usepackage[T2A]{fontenc}
\usepackage[cp1251]{inputenc}
\usepackage{longtable}

\begin{document}

\title{The unusual photometric behavior of the new FUor star V2493 Cyg (HBC 722)}
\titlerunning{The unusual photometric behavior of the new FUor star V2493 Cyg (HBC 722)}
\author{Evgeni H. Semkov\inst{1}, Stoyanka P. Peneva\inst{1}, Sunay I. Ibryamov\inst{1}, Dinko P. Dimitrov\inst{1}}
\authorrunning{E. Semkov et al.}
\tocauthor{Evgeni Semkov}
\institute{Institute of Astronomy and National Astronomical Observatory, Bulgarian Academy of Sciences, Sofia, Bulgaria
        \newline
        \email{esemkov@astro.bas.bg}
}
\papertype{Research report. Accepted on xx.xx.xxxx}
\maketitle

\begin{abstract}
The recent results from photometric study of the new FUor star found in the field of NGC 7000/IC 5070 are presented in the paper. 
The outburst of V2493 Cyg in the summer of 2010 generated considerable interest among the astronomical community.
V2493 Cyg is the first FUor object, whose outburst was observed from its very beginning in all spectral ranges.
After reaching the firs maximum in September/October 2010, the brightness of V2493 Cyg declined slowly, having weakened by 1\fm45 ($V$) by the spring/early summer of 2011.
Since the autumn of 2011, another light increase occurred and the star became brighter by 1\fm8 ($V$) until April 2013. 
The recent photometric data show that the star keeps its maximum brightness during the period April - August 2013 and the recorded amplitude of the outburst reaches $\Delta$$V$=5\fm1.
Consequently, the outburst of V2493 Cyg lasts for more than three years. 
We expect that the interest in this object will increase in the coming years and the results will help to explore the nature of young stars.
\end{abstract}

\keywords{Stars: pre-main sequence, Stars: variables: T Tauri, FU Orionis, Stars: individual: V2493 Cyg}

\section{Introduction}

The study of the large amplitude brightness variability of pre-main sequence (PMS) stars is of great importance in understanding stellar evolution. 
These variations comprise transient increases in brightness (outbursts), temporary drops in brightness (eclipses), and large amplitude irregular or regular variations for a short or long time scales.

One of the most remarkable phenomenon in the early stages of stellar evolution is the FU Orionis (FUor) outbursts.
The flare-up of FU Orionis itself was documented by Wachmann (1939) and for several decades it was the only known object of that type. 
Herbig (1977) defined FUors as a class of young variables after the discovery of outbursts from two new objects - V1057 Cyg and V1515 Cyg. 
The main characteristics of FUors are an increase in optical brightness of about 4-5 mag, a F-G supergiant spectrum with broad blue-shifted Balmer lines, strong infrared excess, connection with reflection nebulae, and location in star-forming regions (Reipurth \& Aspin 2010). 
Typical spectroscopic properties of FUors include a gradual change in the spectrum from earlier to later spectral type from the blue to the infrared, a strong Li I ($\lambda$ 6707) line, P Cygni profiles of H$\alpha$ and Na I ($\lambda$ 5890/5896) lines, and the presence of CO bands in the near infrared spectra (Herbig 1977, Bastian \& Mundt 1985). 
The light curves of FUors are varying from one object to another, but in the most cases the rise goes faster than decline in brightness.
A typical outburst of FUor objects can last for several decades or even a century.
Registration of an outburst in the optical wavelengths is considered to be a necessary condition a single PMS star to be accepted as FUor.
Therefore, any new announcement for registration of an eruption from PMS star is welcomed with a great interest by researchers.

FUor stars seem to be related to the low-mass PMS objects (T Tauri stars), which have massive circumstellar disks. 
The widespread explanation of the FUor phenomenon is a sizable increase in the disc accretion rate onto the stellar surface.
The cause of increased accretion appears to be thermal or gravitational instability in the circumstellar disk (Hartmann \& Kenyon 1996).
During the outburst, accretion rates raze from $\sim$10$^{-7}$$M_{\sun}$$/$yr to $\sim$10$^{-4}$$M_{\sun}$$/$yr which changes significantly the circumstellar environment. 
The surface temperature of the disk becomes 6000-8000 $K$ and it radiates most of its energy in the optical wavelengths.
For the period of $\sim$100 years the circumstellar disk adds $\sim$10$^{-2}M_{\sun}$ onto the central star and it ejects $\sim$10\% of the accreting material in a high velocity stellar wind. 
Some FUor objects were found to exhibit periodic spectroscopic (Herbig et al. 2003, Powell et al. 2013) or low amplitude photometric (Kenyon et al. 2000, Green et al. 2013, Siwak et al. 2013) variability in short time-scale (days).

On the summer of 2010 we discovered a large amplitude outburst from a PMS star (Semkov \& Peneva 2010a) located in the dark clouds (so-called "Gulf of Mexico") between NGC 7000 (the North America Nebula) and IC 5070 (the Pelican Nebula). 
The outburst was independently discovered by Miller et al. (2011) during the regular monitoring of NGC 7000 with the Palomar 48-in telescope.
The outburst of V2493 Cyg generated considerable interest and was studied across a wide spectral range.
Follow-up photometric observations by Semkov \& Peneva (2010b, 2011), Miller et al. (2011), Leoni et al. (2010), Semkov et al. (2010, 2012), Lorenzetti et al. (2011), and K{\'o}sp{\'a}l et al. (2011) recorded an ongoing light increase in both the optical and infrared.
Follow-up high and low resolution spectroscopic observations by Munari et al. (2010), Miller et al. (2011), Lee et al. (2011), Lorenzetti et al. (2012) and Semkov et al. (2010, 2012) showed significant changes in both the profiles and intensity of the spectral lines.
The pre-outburst spectral energy distribution (SED) of V2493 Cyg is discussed in the papers of Miller et al. (2011) and K{\'o}sp{\'a}l et al. (2011). 
The authors concluded that before the eruption V2493 Cyg was a Class II young stellar object - most often associated with Classical T Tauri stars.

After reaching its maximum brightness in October 2010, the brightness of V2493 Cyg declined slowly, having weakened by 1\fm45 ($V$) by the spring/early summer of 2011 (Semkov et al. 2012, Lorenzetti et al. 2012).
Since the autumn of 2011, another light increase occurred, which continued in 2012 and 2013 (Semkov et al. 2012, Antoniucci et al. 2013).

\section{Observations}

The present paper is a continuation of our photometric study of V2493 Cyg during the outburst (Semkov et al. 2010, 2012).
We present new $BVRI$ photometric data of the star in the period May 2012 - August 2013.
The CCD photometric observations of V2493 Cyg were performed with the 2-m RCC, the 50/70-cm Schmidt, and the 60-cm Cassegrain telescopes of the National Astronomical Observatory Rozhen (Bulgaria) and the 1.3-m RC telescope of the Skinakas Observatory of the Institute of Astronomy, University of Crete (Greece).
Observations were performed with four types of the CCD camera — Vers Array 1300B at the 2-m RCC telescope, ANDOR DZ436-BV at the 1.3-m RC telescope, FLI PL16803 at the 50/70-cm Schmidt telescope, and FLI PL9000 at the 60-cm Cassegrain telescope.

All the data were analyzed using the same aperture, which was chosen as 4\arcsec  in radius (while the background annulus was from 13\arcsec to 19\arcsec) in order to minimize the light from the surrounding nebula and avoid contamination from adjacent stars. 
As references, we used the $BVRI$ comparison sequence of fifteen stars in the field around V2493 Cyg published in Semkov et al. (2010). 
In this way we provided a maximum consistency of the photometric results obtained at different telescopes and CCD cameras.
The results of our photometric CCD observations of V2493 Cyg are summarized in Table 1.  
The columns provide the date and Julian date (JD) of observation, $\it IRVB $ magnitudes of V2493 Cyg, the telescope and CCD camera used. 
The typical errors in the reported magnitudes are $0\fm01$ for $I$ and $R$-band data, $0\fm01$-$0\fm02$ for $V$, and $0\fm02-0\fm05$ for $B$-band.

The construction of the historical light curve of V2493 Cyg is important for the determination of the outburst mechanism.
The only one possible way to study the long-term variability of PMS stars is exploration of the photographic plate archives.
The bright North America and Pelican nebulae have been continuously attracting the interest of astrophotographers and
researchers worldwide, and as a consequence the plate archives of several observatories preserve abundant collections of plates exposed over several
decades on this region of the sky.
The collection and analysis of all these observations is very valuable for the study of the long-time variability of V2493 Cyg but it requires a very long and laborious amount of work.
Until now we have obtained the results of exploring the whole photographic plate stack preserved at two observatories, the Asiago Observatory (Italy), and the National Astronomical Observatory Rozhen (Bulgaria).
Photographic observations in the field of V2493 Cyg were performed with the 67/92 cm and the 40/50 cm Schmidt telescopes at the Asiago Observatory
and with the 2-m RCC telescope and the 50/70 cm Schmidt telescope at the Rozhen Observatory.
The digitized plates from the Palomar Schmidt telescope, available via the website of the Space Telescope Science Institute, are also used.
We also used several scanned plates from the 100/130 cm Schmidt telescope of the Byurakan Astrophysical Observatory (Armenia), the 30 cm Astrograph of the Hoher List Observatory (Germany), and the historic first Schmidt-type telescope (36/44 cm) mounted on the Hamburg-Bergedorf Observatory (Germany) (Semkov et al. 2012). 

\begin{longtable}{llllllll}
\caption{Photometric CCD observations of V2493 Cyg}\\
\hline\hline
\noalign{\smallskip}  
Date \hspace{1.5cm} &	J.D. (24...) \hspace{2mm}	&	I	\hspace{8mm} & R \hspace{8mm} & V \hspace{8mm} & B \hspace{8mm} & Telescope \hspace{1mm} & CCD	\\
\noalign{\smallskip}  
\hline
\endfirsthead
\caption{continued.}\\
\hline\hline
\noalign{\smallskip}  
Date \hspace{1.5cm} &	J.D. (24...) \hspace{2mm}	&	I	\hspace{8mm} & R \hspace{8mm} & V \hspace{8mm} & B \hspace{8mm} & Telescope \hspace{1mm} & CCD	\\
\noalign{\smallskip}  
\hline
\noalign{\smallskip}  
\endhead
\hline
\endfoot
\noalign{\smallskip}
2012 May 12 & 56060.415 & 11.78 & 12.91 & 13.99 & 15.55 & Sch  & FLI\\
2012 May 20 & 56068.399 & 11.83 & 12.97 & 14.03 & 15.48 & Sch  & FLI\\
2012 Jun 12 & 56091.398 & 11.76 & 12.94 & 14.00 & 15.54 & Sch  & FLI\\
2012 Jun 13 & 56092.383 & 11.75 & 12.92 & 14.00 & 15.54 & Sch  & FLI\\
2012 Jun 15 & 56094.441 & 11.66 & 12.81 & 13.85 & 15.42 & 2m   & VA\\
2012 Jun 17 & 56096.398 & 11.66 & 12.82 & 13.88 & 15.42 & Sch  & FLI\\
2012 Jul 11 & 56120.377 & 11.69 & 12.85 & 13.91 & 15.45 & Sch  & FLI\\
2012 Jul 12 & 56121.338 & 11.67 & 12.84 & 13.89 & 15.43 & Sch  & FLI\\
2012 Jul 13 & 56122.399 & 11.64 & 12.80 & 13.86 & 15.37 & Sch  & FLI\\
2012 Jul 14 & 56123.387 & 11.65 & 12.80 & 13.87 & 15.40 & Sch  & FLI\\
2012 Jul 30 & 56139.279 & 11.62 & 12.78 & 13.84 & 15.37 & 1.3m & ANDOR\\
2012 Aug 01 & 56141.416 & 11.63 & 12.80 & 13.86 & 15.39 & 1.3m & ANDOR\\
2012 Aug 02 & 56142.270 & 11.62 & 12.77 & 13.83 & 15.37 & 1.3m & ANDOR\\
2012 Aug 03 & 56143.254 & 11.56 & 12.71 & 13.77 & 15.32 & 1.3m & ANDOR\\
2012 Aug 04 & 56144.251 & 11.59 & 12.74 & 13.81 & 15.35 & 1.3m & ANDOR\\
2012 Aug 11 & 56150.615 & 11.56 & 12.72 & 13.81 & 15.35 & 1.3m & ANDOR\\
2012 Aug 12 & 56151.621 & 11.62 & 12.77 & 13.81 & 15.35 & 1.3m & ANDOR\\
2012 Aug 13 & 56152.624 & 11.60 & 12.76 & 13.82 & 15.35 & 1.3m & ANDOR\\
2012 Aug 14 & 56153.624 & 11.55 & 12.71 & 13.77 & 15.32 & 1.3m & ANDOR\\
2012 Aug 15 & 56154.622 & 11.55 & 12.73 & 13.79 & 15.34 & 1.3m & ANDOR\\
2012 Aug 16 & 56155.620 & 11.57 & 12.74 & 13.79 & 15.32 & 1.3m & ANDOR\\
2012 Aug 16 & 56156.242 & 11.59 & $-$   & $-$   & $-$   & 1.3m & ANDOR\\
2012 Aug 18 & 56157.582 & 11.54 & 12.68 & 13.75 & 15.29 & 1.3m & ANDOR\\
2012 Aug 19 & 56159.351 & 11.50 & 12.63 & 13.68 & 15.20 & Sch  & FLI\\
2012 Aug 20 & 56160.329 & 11.56 & 12.70 & 13.76 & 15.29 & Sch  & FLI\\
2012 Aug 21 & 56160.529 & 11.54 & 12.71 & 13.79 & 15.34 & 1.3m & ANDOR\\
2012 Aug 21 & 56161.346 & 11.53 & 12.67 & 13.73 & 15.23 & Sch  & FLI\\
2012 Aug 22 & 56162.338 & 11.53 & 12.68 & 13.74 & 15.26 & Sch  & FLI\\
2012 Aug 29 & 56169.275 & 11.56 & 12.72 & 13.77 & 15.31 & Sch  & FLI\\
2012 Sep 02 & 56173.335 & 11.57 & 12.75 & 13.82 & 15.36 & 1.3m & ANDOR\\
2012 Sep 03 & 56174.306 & 11.58 & 12.76 & 13.83 & 15.39 & 1.3m & ANDOR\\
2012 Sep 07 & 56178.259 & 11.64 & 12.83 & 13.90 & 15.44 & 1.3m & ANDOR\\
2012 Sep 08 & 56179.467 & 11.61 & 12.77 & 13.83 & $-$   & 1.3m & ANDOR\\
2012 Sep 09 & 56180.312 & 11.60 & 12.77 & 13.83 & 15.37 & 1.3m & ANDOR\\
2012 Sep 10 & 56181.288 & 11.57 & 12.74 & 13.81 & 15.36 & 1.3m & ANDOR\\
2012 Sep 11 & 56182.246 & 11.53 & 12.70 & 13.76 & 15.31 & 1.3m & ANDOR\\
2012 Sep 12 & 56183.370 & 11.56 & 12.73 & 13.79 & 15.35 & 1.3m & ANDOR\\
2012 Sep 22 & 56193.282 & 11.55 & 12.73 & 13.80 & 15.35 & 1.3m & ANDOR\\
2012 Sep 22 & 56193.339 & 11.55 & 12.72 & 13.78 & 15.32 & Sch  & FLI\\
2012 Sep 23 & 56194.322 & 11.52 & 12.67 & 13.71 & 15.25 & Sch  & FLI\\
2012 Sep 24 & 56195.250 & 11.48 & 12.62 & 13.66 & 15.22 & Sch  & FLI\\
2012 Oct 07 & 56208.210 & 11.53 & 12.66 & 13.70 & 15.24 & Sch  & FLI\\
2012 Oct 08 & 56209.204 & 11.57 & 12.71 & 13.75 & 15.27 & Sch  & FLI\\
2012 Oct 09 & 56210.193 & 11.56 & 12.70 & 13.74 & 15.26 & Sch  & FLI\\
2012 Oct 10 & 56211.461 & 11.50 & 12.63 & 13.67 & $-$   & Sch  & FLI\\
2012 Oct 11 & 56212.261 & 11.47 & 12.59 & 13.64 & 15.15 & 60cm & FLI\\
2012 Oct 13 & 56214.193 & 11.51 & 12.64 & 13.68 & 15.20 & 2m   & VA\\
2012 Oct 25 & 56226.240 & 11.40 & 12.55 & 13.61 & 15.11 & Sch  & FLI\\
2012 Oct 26 & 56227.401 & 11.48 & 12.63 & 13.70 & 15.24 & Sch  & FLI\\
2012 Nov 17 & 56249.172 & 11.41 & 12.55 & 13.60 & 15.13 & Sch  & FLI\\
2012 Nov 18 & 56250.181 & 11.44 & 12.58 & 13.64 & 15.18 & Sch  & FLI\\
2012 Dec 12 & 56274.165 & 11.47 & 12.61 & 13.67 & 15.20 & 2m   & VA\\
2012 Dec 13 & 56275.217 & 11.53 & 12.66 & 13.74 & 15.26 & 2m   & VA\\
2012 Dec 14 & 56276.166 & 11.51 & 12.61 & 13.69 & 15.22 & 2m   & VA\\
2012 Dec 15 & 56277.163 & 11.52 & 12.61 & 13.65 & $-$   & 2m   & VA\\
2012 Dec 31 & 56293.261 & 11.46 & 12.61 & 13.67 & 15.20 & Sch  & FLI\\
2013 Jan 01 & 56294.236 & 11.41 & 12.56 & 13.63 & 15.19 & 60cm & FLI\\
2013 Jan 03 & 56296.270 & 11.39 & 12.51 & 13.56 & $-$   & 60cm & FLI\\
2013 Jan 16 & 56309.238 & 11.36 & 12.50 & 13.54 & 15.09 & Sch  & FLI\\
2013 Jan 19 & 56312.227 & 11.33 & 12.42 & 13.47 & 15.01 & 2m   & VA\\
2013 Feb 04 & 56328.200 & 11.23 & 12.36 & 13.41 & $-$   & Sch  & FLI\\
2013 Feb 05 & 56329.191 & 11.22 & 12.35 & 13.38 & 14.91 & Sch  & FLI\\
2013 Mar 06 & 56357.590 & 11.20 & 12.29 & 13.34 & 14.90 & 60cm & FLI\\
2013 Mar 18 & 56369.610 & 11.16 & 12.25 & 13.30 & 14.85 & 2m   & VA\\
2013 Apr 10 & 56392.511 & 11.12 & 12.22 & 13.27 & 14.71 & Sch  & FLI\\
2013 Apr 11 & 56394.482 & 11.15 & 12.27 & 13.30 & 14.80 & Sch  & FLI\\
2013 May 02 & 56415.419 & 11.15 & 12.25 & 13.29 & 14.81 & Sch  & FLI\\
2013 May 04 & 56417.468 & 11.13 & 12.20 & 13.25 & 14.82 & 2m   & VA\\
2013 May 15 & 56428.405 & 11.17 & 12.28 & 13.33 & 14.86 & 60cm & FLI\\
2013 May 17 & 56430.408 & 11.20 & 12.31 & 13.36 & 14.87 & 60cm & FLI\\
2013 May 19 & 56432.406 & 11.16 & 12.26 & 13.29 & 14.74 & 60cm & FLI\\
2013 May 30 & 56443.357 & 11.21 & 12.33 & 13.36 & 14.87 & Sch  & FLI\\
2013 May 31 & 56444.355 & 11.15 & 12.28 & 13.33 & 14.80 & Sch  & FLI\\
2013 Jul 04 & 56478.364 & 11.19 & 12.30 & 13.35 & 14.87 & 2m   & VA\\
2013 Aug 01 & 56506.283 & 11.21 & 12.33 & 13.39 & 14.94 & 2m   & VA\\
2013 Aug 02 & 56507.273 & 11.22 & 12.34 & 13.40 & 14.93 & 2m   & VA\\
2013 Aug 03 & 56508.319 & 11.19 & 12.32 & 13.36 & 14.92 & 2m   & VA\\
2013 Aug 04 & 56509.289 & 11.11 & 12.23 & 13.29 & 14.80 & Sch  & FLI\\
2013 Aug 05 & 56510.370 & 11.09 & 12.21 & 13.27 & 14.75 & 60cm & FLI\\
2013 Aug 05 & 56510.381 & 11.09 & 12.21 & 13.28 & 14.79 & Sch  & FLI\\
2013 Aug 06 & 56511.412 & 11.09 & 12.23 & 13.26 & 14.78 & Sch  & FLI\\
2013 Aug 06 & 56511.413 & 11.08 & 12.21 & 13.28 & 14.76 & 60cm & FLI\\
2013 Aug 07 & 56512.399 & 11.14 & 12.27 & 13.31 & 14.84 & Sch  & FLI\\
2013 Aug 07 & 56512.404 & 11.11 & 12.25 & 13.30 & 14.85 & 60cm & FLI\\
2013 Aug 08 & 56513.382 & 11.13 & 12.29 & 13.35 & 14.90 & 60cm & FLI\\
2013 Aug 09 & 56514.350 & 11.12 & 12.28 & 13.33 & 14.80 & 60cm & FLI\\
2013 Aug 12 & 56517.284 & 11.11 & 12.26 & 13.31 & 14.86 & 60cm & FLI\\
\end{longtable}
 
\section{Results}

The $BVRI$ light-curves of V2493 Cyg during the period September 1973 $-$ August 2013 are plotted in Fig. 1.
The filled diamonds represent our CCD observations (Semkov et al. 2010, 2012 and the present paper),
the filled circles observations from the 48 inch Samuel Oschin telescope at Palomar Observatory (Miller et al. 2011), 
the open diamonds photographic data from the Asiago Schmidt telescopes,
the open squares photographic data from the Palomar Schmidt telescope,
the filled squares photographic data from the Byurakan Schmidt telescope
and the open circles photographic data from the Rozhen 2-m RCC telescope.
The optical photometric data published by other authors (K{\'o}sp{\'a}l et al. 2011; Lorenzetti et al. 2012) show the same changes in brightness during the outburst, which are perfectly matched with the light curves shown in Fig 1.
  
The photometric observation obtained before the outburst displayed only small amplitude variations in all pass-bands typical of T Tauri stars.  
Other large-amplitude eruptions have not been registered in our long-term photometric study.
The observational data indicate that the outburst started sometime before May 2010, and reached the first maximum value in September/October 2010.  
Since October 2010, a slow fading was observed and up to May 2011 the star brightness decreased by 1\fm4 ($V$). 
In the period from May till October 2011 no significant changes in the brightness of the star are observed, its brightness remains at 3\fm3 ($V$) above the quiescence level.
Since the autumn of 2011, another light increase occurred and the star became brighter by 1\fm8 ($V$) until April 2013. 
During the period April - August 2013 the star keeps its maximum brightness showing a little bit fluctuations around it.
By comparing with brightness levels in 2009, we derive the following values for the outburst amplitude: $\Delta$$I$=4\fm1, $\Delta$$R$=4$\fm$7, $\Delta$$V$=5\fm1, and $\Delta$$B$=5\fm1.  
Simultaneously with the increase in brightness the star color changes significantly, the star become appreciably bluer.

\begin{figure}[!htb]
  \begin{center}
   \centering{\epsfig{file=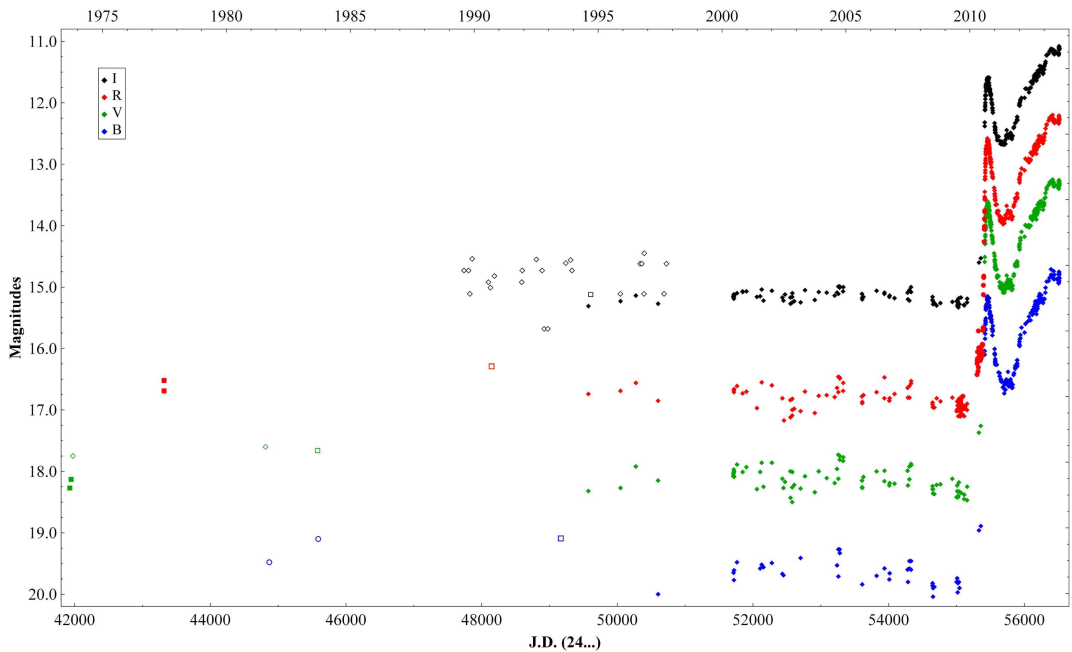}}
    \caption[]{$BVRI$ light curves of V2493 Cyg for the period Sep. 1973 - Aug. 2013}
    \label{countryshape}
  \end{center}
\end{figure}

Simultaneously with the optical outburst appearance of a reflection nebula around V2493 Cyg was observed.
Such reflection nebulae are typical of all classical FUor objects.
The PMS stars are located in regions of star formation along with large dark clouds of interstellar medium.
During the FUor type outburst the brightness of V2493 Cyg has increased by about hundred times and this makes visible a part of the dark matter around the star. 
In Fig. 2 we show color images of the area around V2493 Cyg, obtained in three different epochs $-$ before the outburst, at the time of the first maximum and during the second maximum in brightness.
The dates of the images shown were selected from nights with a relatively very good atmospheric seeing $-$ around 1 arc second.
The relatively bright nebula around the star has not changed significantly between the two observed maximums in brightness.

\begin{figure}[!htb]
  \begin{center}
   \centering{\epsfig{file=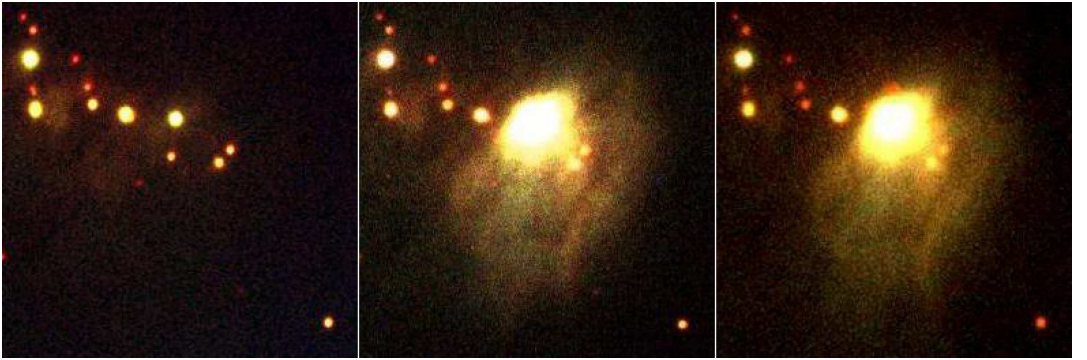}}
    \caption[]{Color images of V2493 Cyg obtained with the 2-m RCC telescope in NAO Rozhen {\it Left}: on 2007 Aug. 16, 
   {\it Center}: on 2010 Oct. 31, {\it Right}: on 2013 Aug. 03}
    \label{countryshape}
  \end{center}
\end{figure}

Fig. 3 represents the change in position of V2493 Cyg during the outburst on the color-color diagram $V-R/R-I$.
The location of main-sequence dwarfs (blue line) according to Johnson (1966) and the interstellar reddening vector for the region A$_{V}$ = 3\fm4 (black line) according to Cohen \& Kuhi (1979) are shown. 
Data for the main-sequence stars are recalculated to correspond from the Johnson's to Cousins system using the corresponding equations in Moro \& Munari (2000).
Fig. 3 shows that V2493 Cyg becomes appreciably bluer when its brightness increases during the outburst.  

\begin{figure}[!htb]
  \begin{center}
   \centering{\epsfig{file=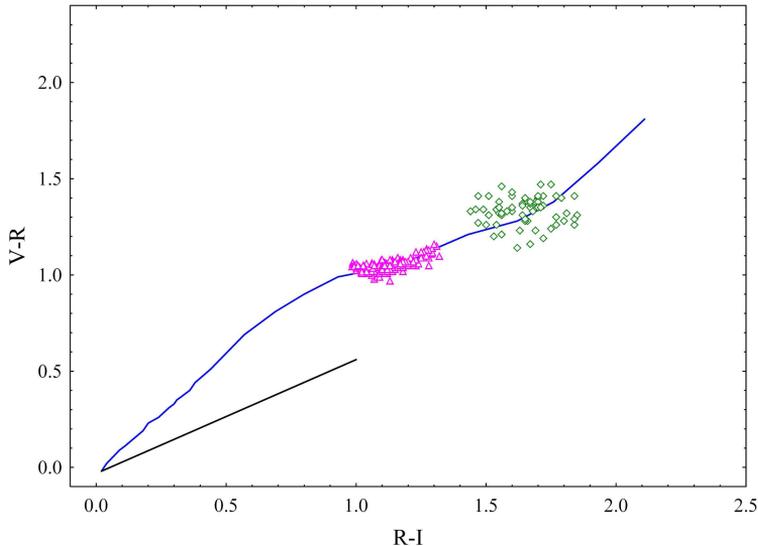}}
    \caption[]{$V-R$ vs. $R-I$ color-color diagram for V2493 Cyg before (open diamonds) and during the outburst (open triangles). The solid blue line
is the locus of main-sequence dwarf stars. The straight black line is the reddening vector for $A_{V}$ = 3\fm4.}
    \label{countryshape}
  \end{center}
\end{figure}

\section{Discussion and conclusion}

During the three years passing from the beginning of the outburst of V2493 Cyg the photometric data show interesting and somewhat unexpected results. 
From all objects associated with the group of FUors, only three (FU Ori, V 1057 Cyg, and V 1515 Cyg) have detailed photometric observations taken during the outburst and during the set of brightness (Clarke et al. 2005).
Because of the wide interest from photometric observations V2493 Cyg will be the fourth such object with a well studied light curve.
For some years, we have made efforts to construct the historical light curves of several other FUor and FUor-like objects such as: V1735 Cyg (Peneva et al. 2009), Parsamian 21 (Semkov \& Peneva 2010c), V733 Cep (Peneva et al. 2010), V1647 Ori (Garc\'ia-Alvarez et al. 2011), and V582 Aur (Semkov et al. 2013).
To realize this study, we use both data from recent CCD photometric monitoring as well as photometric data from the photographic plate archives.
Our results suggest that each object of FUor type has a characteristic long-term light curve, which distinguishes it from other objects. 
The shape of the observed light curves of FUors may vary considerably in the time of rise, the rate of decrease in brightness, the time spent at maximum brightness, and the light variability during the set in brightness. 

The light curve of V2493 Cyg from all available photometric observations is also somewhat unique.
The rate of increase in the brightness (the fastest ever recorded) was followed by a very rapid fall in brightness.
During the period of rise in brightness and the first months after the maximum, the light curve of V2394 Cyg is similar to the light curves of the classical FUor object V1057 Cyg and FU Ori itself.
But the most remarkable feature of the light curve of V2493 Cyg is the repeated rise in brightness in the past 1.5 years and the reaching of a second maximum in brightness. 

The observed double maximum and the large amplitude of brightness between the two peaks may result from a variable accretion from the circumstellar disk onto the stellar surface.
One of the possible reasons for the variable accretion rate could be fragmentation of the circumstellar disk.
Because the FUor phenomenon is probably repeatable, up to 50\% of the protostellar mass can be accumulated as a result of such episodes of strong accretion burst.
Stamatellos et al. (2012) suggest that episodic accretion may initially promote disc fragmentation. 
In the early stages of PMS evolution, fragmentation does not happen and disk accretion is assumed to be constant.
After several episodic accretion bursts, the circumstellar disk is gradually fragmented and thus prevents new FUor events.
Therefore, it can be supposed that FUor outbursts during different periods of stellar evolution may vary in amplitude, duration, and shape of the light curve due to the different state of disk fragmentation.
Strong accretion bursts may also be the triggering mechanism of planet formation with different masses inside the circumstellar disk.
If the above suggestions are correct, the studies of FUor objects would be useful not only for understanding stellar evolution but also for understanding the formation of planets and asteroids and the frequency of planetary systems.

We plan to continue our spectroscopic and photometric monitoring of the star during the next few months and years and strongly encourage similar follow-up observations.

{\it Acknowledgments:} This work was partly supported by grant DO 02-85 of the National Science Fund of the Ministry of Education and Science, Bulgaria and by ESF and Bulgarian Ministry of Education and Science under the contract BG051PO001-3.3.06-0047. 
This research has made use of the NASA Astrophysics Data System.

\end{document}